\documentclass[prl,reprint,amssymb]{revtex4-2}

\usepackage{xcolor}

\usepackage{bm}
\usepackage{braket}
\usepackage{amsmath}  
\usepackage{amsfonts}
\usepackage{graphicx}
\usepackage{hyperref}
\hypersetup{
    colorlinks = true,
    citecolor  = blue,
    linkcolor  = red,
    urlcolor   = blue
}

\begin{document}



\title{Optically induced magnetic inertia and magnons from non-Markovian extension of the Landau-Lifshitz-Gilbert equation}


\author{Felipe Reyes-Osorio}
\author{Branislav K. Nikoli\'c}
\email{bnikolic@udel.edu}
\affiliation{Department of Physics and Astronomy, University of Delaware, Newark, DE 19716, USA}


\begin{abstract}
The Landau-Lisfhitz-Gilbert (LLG) equation has been the cornerstone of modeling the dynamics of localized spins, viewed as classical vectors of fixed length, within nonequilibrium magnets. When light is employed as the nonequilibrium drive, the LLG equation must be supplemented with additional terms that are usually conjectured using phenomenological arguments for direct opto-magnetic coupling between localized spins and (real or effective) magnetic field of light. However, direct coupling of magnetic field to spins is $1/c$ smaller than coupling of light and electrons; or both magnetic and electric fields are too fast for slow classical  spins to be able to follow them. Here, we displace the need for phenomenological arguments by rigorously deriving an {\em extended} LLG equation via Schwinger-Keldysh  field theory (SKFT). Within such a theory, light interacts with itinerant electrons, and then  spin current carried by them exerts spin-transfer torque onto localized spins, so that when photoexcited electrons are integrated out we arrive at a spin-only equation. Unlike the standard phenomenological LLG equation with local-in-time Gilbert damping, our extended one contains a {\em non-Markovian} memory kernel whose plot within the plane of its two times arguments exhibits \textit{fractal} properties. By applying  the SKFT-derived  extended LLG equation, as our central result, to a light-driven ferromagnet as an example, we predict an optically induced magnetic inertia term. Its  magnitude is governed by a spatially nonlocal and time-dependent prefactor, leading to the excitation of coherent magnons at sharp frequencies in and outside of the band of incoherent (or thermal) magnons.
\end{abstract}

\maketitle

\textit{Introduction}---The ninety-year-old  Landau-Lifshitz equation~\cite{Landau1935}, with the Gilbert (LLG) form of damping~\cite{Gilbert2004,Saslow2009}, has been  the cornerstone~\cite{Skubic2008,Berkov2008,Evans2014,Kim2010,Vansteenkiste2014,Moreels2024} of modeling dynamics of localized spins within nonequilibrium magnets, whenever such spins can be approximated~\cite{GarciaGaitan2024} by   classical vectors $\mathbf{S}_n(t)$ of  fixed length ($|\mathbf{S}_n(t)|=1$)  localized at sites $n$ of the crystalline lattice. However, since the LLG equation is phenomenological~\cite{Landau1935, Gilbert2004,Saslow2009,Galkina2021} in nature, its applicability to new experimental situations requires additional terms~\cite{Zhang2004a,Krivorotov2007,Satoh2010,Tzschaschel2017,Bertelli2021} that are typically not rigorously derived (as is the case of the LLG equation itself~\cite{Landau1935,Galkina2021}). Instead, they are fitted to experiments with typically only a partial~\cite{Krivorotov2007,Bertelli2021,Satoh2010} success. In the case of magnets driven by laser light of sufficiently low intensity (to avoid demagnetization effects~\cite{Kirilyuk2010,Scheid2022,De2024,KorffSchmising2024} which otherwise change  $|\mathbf{S}_n(t)|$ as nonclassical~\cite{Lee2025,GarciaGaitan2024,GarciaGaitan2025a} effect beyond~\cite{GarciaGaitan2025a,Krieger2015, Dewhurst2018,Kefayati2024, Chen2019a,Chen2019c} capabilities of  LLG description), one finds in the literature arguments for various torque-type terms involving real or optically induced effective magnetic fields, such as the Zeeman torque $\propto \mathbf{B}_L(t) \times \mathbf{S}_n(t)$, where the magnetic field $\mathbf{B}_L(t)$ of light directly couples to localized spins~\cite{Kampfrath2010,Blank2023, Neeraj2021}; torque $\propto \partial_t \mathbf{B}_L(t)\times \mathbf{S}_n(t)$ due to the derivative of the same real magnetic field~\cite{Mondal2019a}; torque due to magnetostriction effect~\cite{Tokura2014, Seifert2019}, where electric field of light $\mathbf{E}_L(t)$ couples to the spin-dependent electric polarization $\propto \mathbf{S}_n(t) \cdot \mathbf{S}_{n+1}(t)$; torque exerted by the angular momentum of light~\cite{Mondal2015, Mondal2016, Mondal2021}, which also provides a contribution to the inverse Faraday effect~\cite{Ziel1965, Pershan1966, Hertel2006, Battiato2014,Tanaka2024,Tzschaschel2017} where torque  $\propto \mathrm{Re}\left[i \mathbf{E}_L(t) \times \mathbf{E}_L^*(t)\right] \times \mathbf{S}_n(t)$ is nonzero for circularly polarized light~\cite{Satoh2010}; and torque from the inverse Cotton-Mouton effect~\cite{Rongione2023, Shen2018,Tzschaschel2017} due to linearly polarized light $\propto g_{\alpha\beta \gamma\delta } S^\beta_n(t) E_L^\gamma(t) E_L^\delta(t)$, where $g_{\alpha \beta\gamma\delta}$ is a symmetric tensor.

However, due to the smallness of $\mathbf{B}_L(t)$ when compared to $\mathbf{E}_L(t)$, direct coupling between light and localized spins is generally smaller than the coupling between light and itinerant electrons by a factor of $1/c$ where $c$ is the speed of light. Thus, in first-principles calculations~\cite{Krieger2015,Dewhurst2018, Chen2019a, Chen2019c,Kefayati2024} via time-dependent density functional theory (TDDFT), which do not make any intuitively motivated assumptions about the underlying physics or the system under investigation, including terms for direct  absorption of light angular momentum by electron spin~\footnote{The terms introduced into TDDFT  calculations to describe  direct coupling of light to electron spin include Zeeman term $-\mathbf{B}\cdot \bm{\sigma}$, or orbital-ion degrees of freedom absorbing the light angular momentum, which then indirectly affects electron spin relaxation via spin-orbit coupling~\cite{Chen2019a, Chen2019c}. Thus, TDDFT calculations of light-driven magnets are virtually always performed~\cite{Krieger2015,Chen2019a, Chen2019c,Kefayati2024}  performed by including {\em only} the vector potential coupled to orbital degrees of freedom of an electron, i.e., via its insertion into the kinetic energy term as we use in our model Hamiltonian [Eqs.~\eqref{eq:hamiltonian_pos} and~\eqref{eq:hamiltonian_mom}] as well.} provides negligible effect~\cite{Chen2019a,Chen2019c}. In addition, slow classical degrees of freedom like $\mathbf{S}_n(t)$ {\em cannot}~\cite{Suresh2023} follow fast oscillations of electromagnetic field of visible or mid-infrared lasers, often employed experimentally~\cite{Kirilyuk2010, Chekhov2021}. Even in THz range~\cite{Kampfrath2010,Blank2023,Chekhov2021, Neeraj2021}, where resonant following is possible in principle, a sufficiently strong laser-spin coupling remains difficult to achieve since the field
amplitude of THz laser pulses is limited compared with visible and mid-infrared lasers. Finally,  experiments~\cite{Zhang2020a,Afanasiev2021,Allington2025,Diederich2025} on optical excitation of magnons in two-dimensional semiconducting magnets, where single- and many-body subgap electronic states can be precisely mapped~\cite{Allington2025} by tuning the laser frequency, reveal how excitation of such electronic states always {\em precedes and mediates} excitation of magnons~\cite{Zhang2020a}. In fact, even in conventional gapless transition metal ferromagnets, recent experimental scrutiny~\cite{KorffSchmising2024} finds that direct and coherent interaction between light and spins~\cite{Dewhurst2018} is of lesser importance than secondary processes initiated by electrons directly interacting with light. 

Therefore, a more realistic physical picture is that of optical spin-transfer torque (STT)~\cite{Nemec2012,Zhang2022a,Kimel2019,FernandezRossier2004a,Nunez2004}  exerted by the spin current of photoexcited electrons onto the localized spins, as itinerant electrons can  instantaneously~\cite{Bajpai2019a,Suresh2023,Kefayati2025} respond to applied laser light. However, constructing the proper STT term in the LLG equation requires spatio-temporal profile of 
spin current~\cite{Ralph2008,Suresh2021} or nonequilibrium spin density~\cite{Joao2024,Petrovic2021,Suresh2021}, where attempts have been made to guess such input quantities  empirically~\cite{Ulrichs2018, Ritzmann2020, Nemec2012,FernandezRossier2004a,Nunez2004} or extract them from experimentally measured quantities~\cite{Jechumtal2024} (which is a  highly ambiguous procedure~\cite{Gorchon2022, Kefayati2025} as spin current is not directly measurable). 

Recently, it has been understood~\cite{ReyesOsorio2024,ReyesOsorio2024b,Verstraten2023,Quarenta2024,Hurst2020,Psaroudaki2017} how to rigorously derive  the LLG equation from nonequilibrium quantum field theory~\cite{Kamenev2023,Gelis2019,Altland2010} in Schwinger-Keldysh formulation. The starting point of such derivations are localized quantum spins interacting with a surrounding bath of electrons~\cite{ReyesOsorio2024,Hurst2020,ReyesOsorio2024b} or phonons~\cite{Verstraten2023, Quarenta2024}. The spin dynamics is then approximated into the classical limit while the bath is integrated out within the functional integral~\cite{Kamenev2023,Gelis2019} of Schwinger-Keldysh field theory (SKFT). The end product of such derivations is not the standard~\cite{Landau1935}, but instead  an {\em extended} LLG equation with a {\em non-Markovian} memory kernel. The kernel encodes time-retardation and damping effects stemming from the bath degrees of freedom that are never infinitely fast and always lag behind~\cite{Bajpai2019, Sayad2015, Bajpai2020} localized spins. 

\begin{figure}
    \centering
    \includegraphics[width=\linewidth]{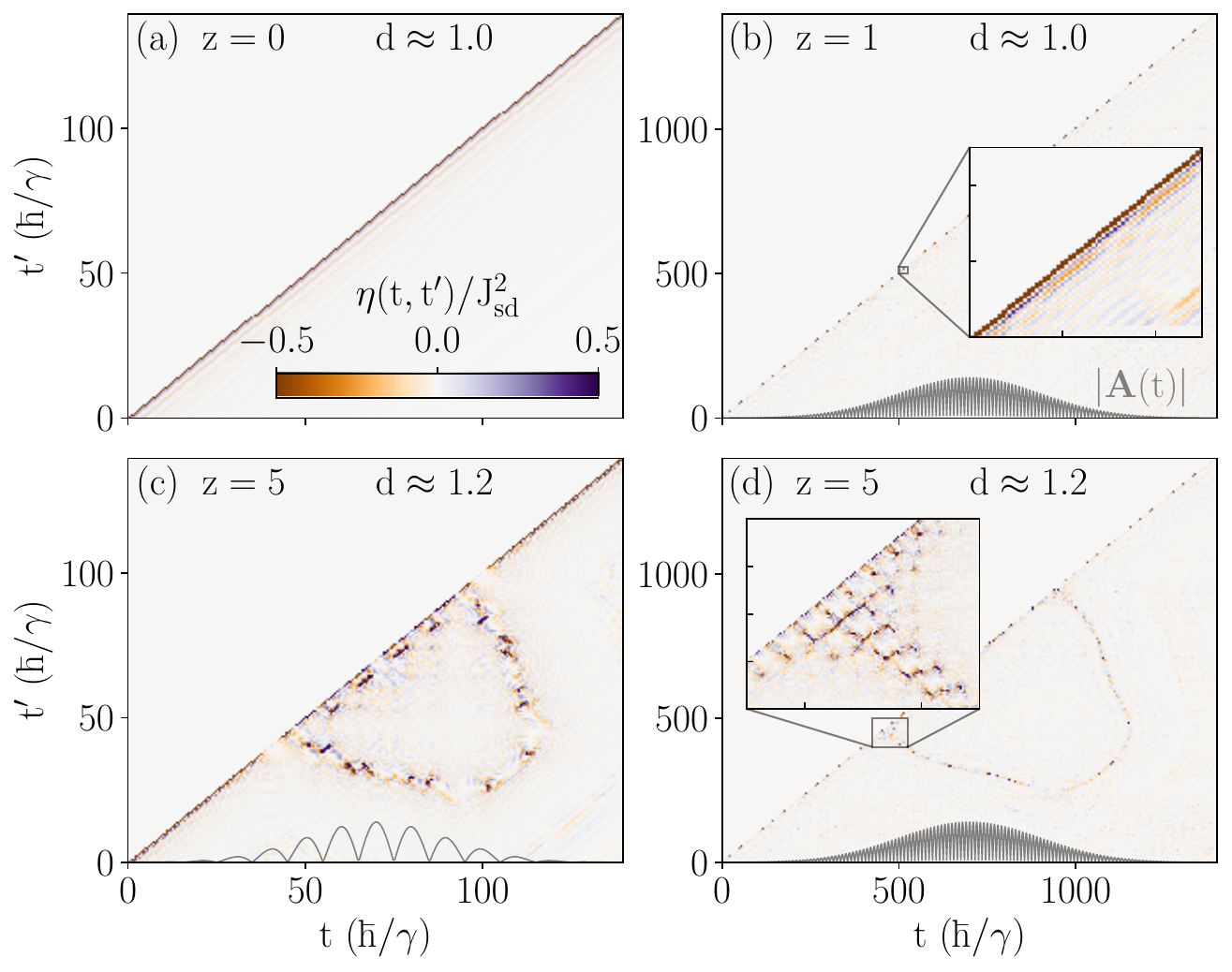}
    \caption{The memory kernel $\eta(t,t')$ [Eq.~\eqref{eq:nonmarkovianKernel}] of SKFT-derived extended LLG Eq.~\eqref{eq:modifiedLLG} plotted as a function of two-times $t$ and $t'$ for an example~\cite{Sayad2015} of a single localized classical spin $\mathbf{S}(t)$ interacting with itinerant electrons in 1D. The kernel is Markovian~\cite{ReyesOsorio2024} in (a), where light is absent; and non-Markovian in (b)--(d), where electrons are driven by fsLP of central frequency $\omega_L=0.3\gamma/\hbar$ and with its vector potential $|\mathbf{A}(t)|$ plotted at the bottom of each panel. The width of fsLP in panels (b) and (d) is $\lambda=200\hbar/\gamma$, whereas in panel (c) it is $\lambda=20\hbar/\gamma$. Finite values of the kernel away from the time diagonal $t=t'$ indicate greater memory effects. As the  amplitude $z$ of fsLP increases in (b)--(d), the plot of $\eta(t,t')$ in $t$-$t'$ plane becomes a {\em fractal} characterized by noninteger dimension $d$.}
    \label{fig:nmkernels}
\end{figure}

\begin{figure}
    \centering
    \includegraphics[width=\columnwidth]{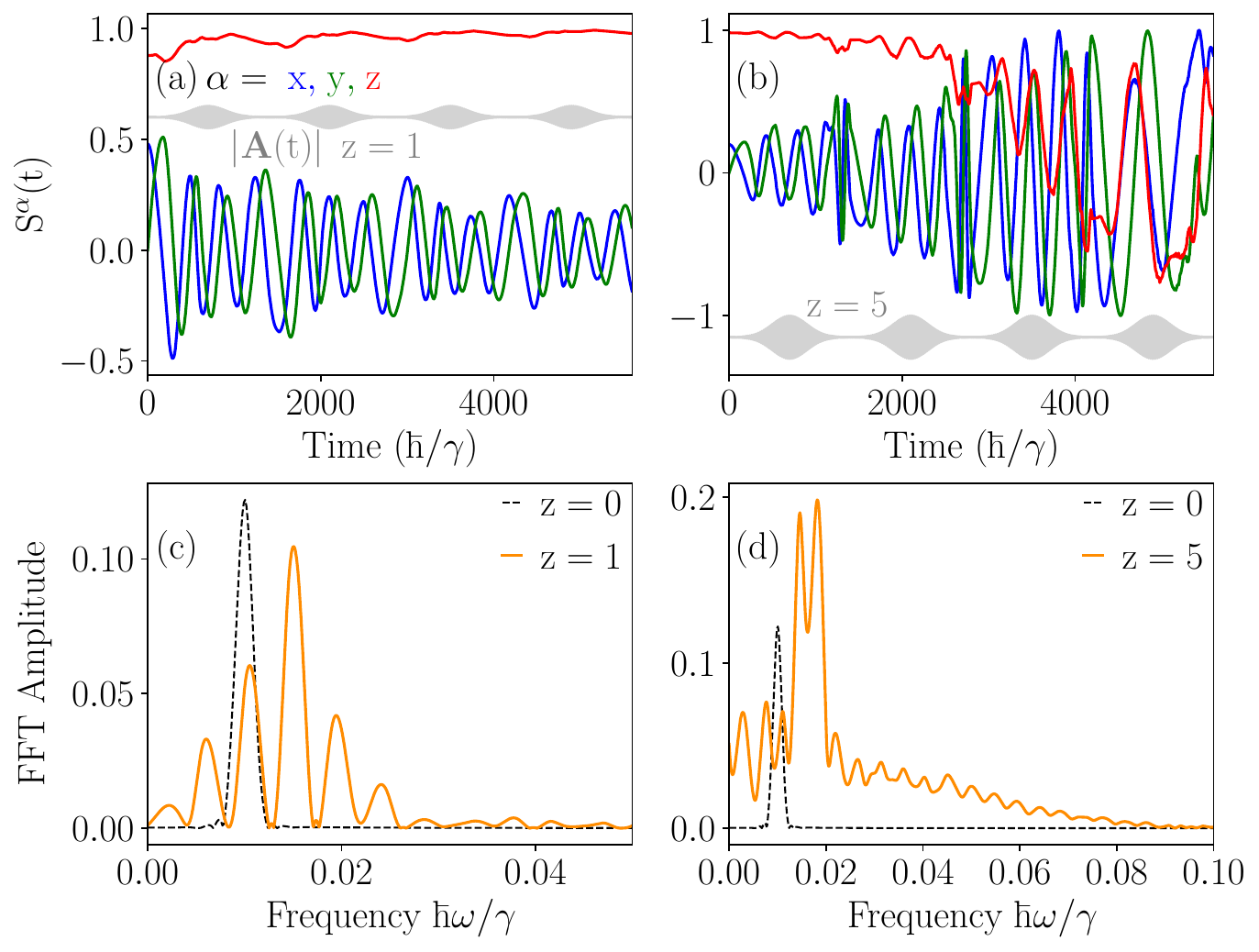}
    \caption{Time evolution of a single localized classical  spin $\mathbf{S}(t)$ interacting with itinerant electrons within an infinite 1D TB chain, as computed from the non-Markovian LLG Eq.~\eqref{eq:modifiedLLG} in an external static magnetic field along the $z$-axis, $\mathbf{B}^{\rm ext}=0.01\gamma \mathbf{e}_z$, and with electrons driven by a train of fsLPs of amplitude (a) $z=1$ and (b) $z=5$, respectively. At $t=0$ spin is slightly tilted away from the $z$-axis, and for its time evolution in (a) and (b) we use the memory  kernels from Figs.~\ref{fig:nmkernels}(b) and~\ref{fig:nmkernels}(d), respectively. Panels (c) and (d) plot FFT amplitude spectra of $S^x(t)$ from panels (a) and (b), respectively. Note that in the absence of fsLPs panel (c) and (d) contain only a single peak (dashed black line) whose position is determined by $|\mathbf{B}^{\rm ext}|$. 
    }
    \label{fig:onespin_nm}
\end{figure}

\begin{figure}
    \centering
    \includegraphics[width=\columnwidth]{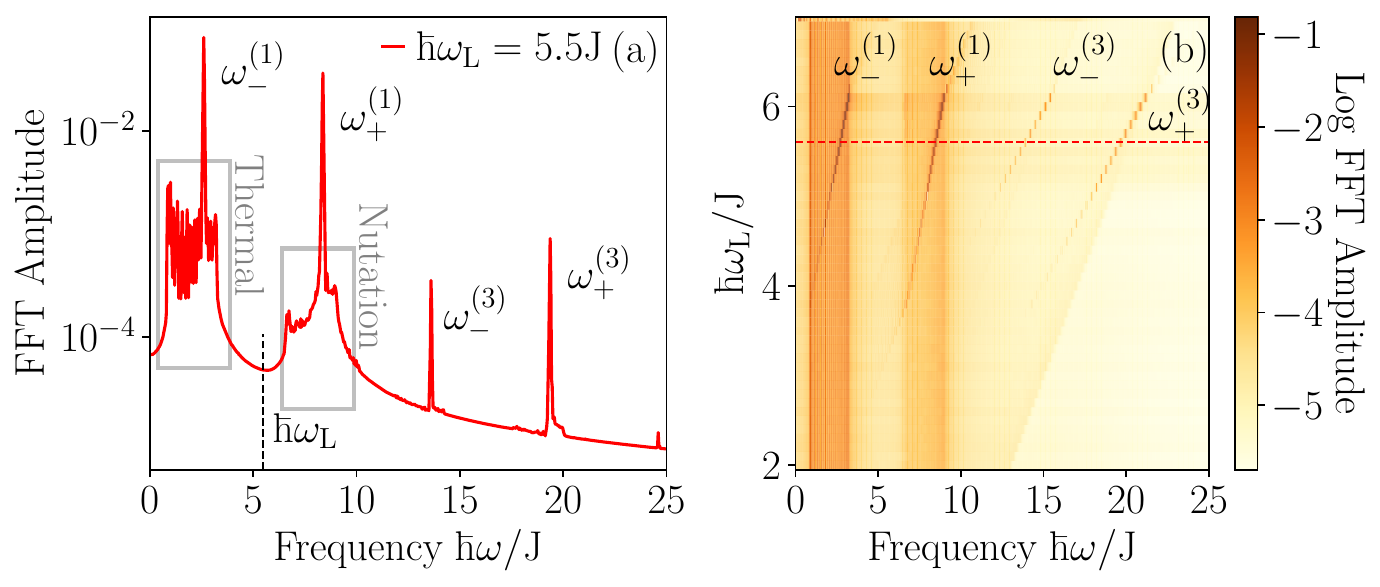}
    \caption{(a) Magnon spectrum of a chain of 30 localized classical spins $\mathbf{S}_n(t)$, as extracted~\cite{Kim2010,Moreels2024}  from FFT of $S^x_{n=15}(t)$ in the middle of the chain, which interact with  itinerant electrons hopping along  an infinite 1D TB chain. The dynamics of $\mathbf{S}_n(t)$, precessing around the $z$-axis, is computed from iLLG Eq.~\eqref{eq:inertia} where the nonequilibrium drive for electrons is CW light of frequency $\omega_L=5.5J/\hbar$ (also marked by vertical dashed line).  The vectors $\mathbf{S}_n(t=0)$ are initially thermalized at \mbox{$k_BT=0.02J$}, so that their evolution via iLLG Eq.~\eqref{eq:inertia} with $I_{nn'}(t)=0$ and $\lambda_{nn'}$ being time-independent~\cite{ReyesOsorio2024} in the absence of light produces a spectrum [enclosed within left gray box in panel (a)] of incoherent (or thermal) magnons~\cite{Pirro2021,De2024}. In contrast, all four sharp peaks in panel (a) at frequencies $\omega^{(1)}_\pm=\omega_L\pm 0.23/\eta_1$ and $\omega^{(3)}_\pm=3\omega_L\pm 0.23/\eta_1$ are optically excited coherent magnons obtained from full iLLG Eq.~\eqref{eq:inertia}. Panel (b) shows FFT amplitude spectra for a range of $\omega_L$, revealing a linear relation between the frequency of optically excited coherent magnon peaks and $\omega_L$. Note that peaks from panel (a) are included in (b) as intersection of dashed red line at $\omega_L=5.5J/\hbar$ and tilted yellow traces.}
    \label{fig:spectra}
\end{figure}

In this Letter, we derive an extended LLG equation via SKFT for a system of localized spins interacting with electrons responding  to laser light, where the latter two are considered as a bath and integrated out. Thus obtained LLG equation 
\begin{align}\label{eq:modifiedLLG}
    \partial_t \mathbf{S}_{n} &= -\mathbf{S}_{n} \times \big(\mathbf{B}^{\rm eff}_{n} +\mathbf{B}^e_{n} - \alpha_G \partial_t \mathbf{S}_{n} \big)  \\
    &+ \mathbf{S}_{n} \times \sum_{n^\prime }\int \! dt' \, \eta_{nn'}(t,t') \mathbf{S}_{n'} (t'), \nonumber
\end{align}
is in general non-Markovian integro-differential one, i.e., dependent on the entire history of the system rather than just its current state; as well as \textit{nonlocal} because it couples spins $\mathbf{S}_n(t)$ and $\mathbf{S}_{n'}(t)$ at any two lattice sites $n$ and $n'$. Here, we use shorthand notation $\partial_t \equiv \partial/\partial t$; $\mathbf{B}^{\rm eff}_n$ is the effective magnetic field at site $n$; $\mathbf{B}^{e}_n$ and $\eta_{nn'}(t,t')$ are the magnetic field and memory kernel, respectively, due to the integrated-out photoexcited electrons; $\alpha_G$ is the dimensionless Gilbert damping parameter~\cite{Gilbert2004, Saslow2009, Evans2014}; and we set the gyromagnetic ratio $g=1$. The non-Markovianity~\cite{Tveten2015} of Eq.~\eqref{eq:modifiedLLG} stems from ultrafast dynamics induced by the driving electric field of light, while its nonlocality~\cite{ReyesOsorio2024} is due to the propagation of electronic spin currents~\cite{Petrovic2021, Bajpai2020, Suresh2021}. When its memory kernel, $\eta_{nn'}(t,t')$, expressed [Eq.~\eqref{eq:nonmarkovianKernel}] in terms of electronic Keldysh Green's functions (GFs)~\cite{Kamenev2023,Stefanucci2013}, is plotted in the plane of two times $t$-$t'$, its geometry exhibits a remarkably complex structure with {\em fractal} properties~\cite{Theiler1990} [Fig.~\ref{fig:nmkernels}(c) and~\ref{fig:nmkernels}(d)] on the proviso that fs laser pulse (fsLP) is  sufficiently intense. For lower fsLP amplitude, $\eta_{nn'}(t,t')$ can be expanded to second order in $t-t'$ to produce an \textit{inertial} LLG (iLLG) equation 
\begin{align}\label{eq:inertia}
    \partial_t \mathbf{S}_{n} &= -\mathbf{S}_{n} \times \big(\mathbf{B}^{\rm eff}_{n} +\mathbf{B}^e_{n} - \alpha_G \partial_t \mathbf{S}_{n} \big)  \\
    &+ \mathbf{S}_{n} \times \sum_{n^\prime} \left [\lambda_{nn'}(t) \partial_t \mathbf{S}_{n'} + I_{nn'}(t) \partial_t^2 \mathbf{S}_{n'} \right],  \nonumber
\end{align}
where magnetic inertia~\cite{Ciornei2011, Faehnle2011, Mondal2017, Thonig2017,  Bajpai2019, Quarenta2024, Mondal2023, Li2015,  Neeraj2021, Kikuchi2015, Daquino2023, Titov2022, Lomonosov2021, Mondal2022, Unikandanunni2022, Rodriguez2024,De2025} is the last term on the right-hand side.  Magnetic inertia and its physical origins have been the subject of recent intense theoretical~\cite{Ciornei2011, Faehnle2011, Mondal2017, Thonig2017,  Bajpai2019, Quarenta2024, Mondal2023}, experimental~\cite{Li2015, Neeraj2021, Unikandanunni2022,De2025} and applications~\cite{Rodriguez2024} pursuits  because of its ``neglect'' in the original LLG equation~\cite{Landau1935, Saslow2009,Gilbert2004,Galkina2021} where only $\propto \mathbf{S}_{n}, \ \partial_t \mathbf{S}_{n}$ terms are included. Unlike recently derived magnetic inertia terms with constant prefactor---such as, due to relativistic effects~\cite{Mondal2017,Thonig2017}, phonon bath~\cite{Quarenta2024} or outflowing spin angular momentum from open systems~\cite{Bajpai2019}---the one driven by light in Eq.~\eqref{eq:inertia} contains {\em time-dependent} prefactor $I_{nn'}(t)$ which is also {spatially nonlocal}. Note that spatial nonlocality of Gilbert damping prefactor $\lambda_{nn'}(t)$ in iLLG Eq.~\eqref{eq:inertia} has been found in prior SKFT derivations~\cite{ReyesOsorio2024}, but its time-dependence in this study is a consequence of light. The non-Markovian LLG  Eq.~\eqref{eq:modifiedLLG} and its approximation in the form of iLLG Eq.~\eqref{eq:inertia} are the  \textit{central results} of our study, whose predictions are illustrated on examples in Figs.~\ref{fig:onespin_nm} and~\ref{fig:spectra} where a single or many, respectively, slow classical localized spins $\mathbf{S}_n(t)$ interact with itinerant  electrons responding fast to light acting as their nonequilibrium drive. The details of our Hamiltonian model and SKFT methodology, leading to Eqs.~\eqref{eq:modifiedLLG} and~\eqref{eq:inertia}, are provided in the End Matter.


\textit{Results and Discussion}---To gain insight into the effects of the non-Markovian nature of SKFT-derived LLG Eq.~\eqref{eq:modifiedLLG}, we first apply it to a simple model of a single localized spin $\mathbf{S}(t)$ which interacts via \textit{sd} exchange of strength $J_{sd}$~\cite{Cooper1967, Ralph2008} with spin of itinerant electrons within an infinite one-dimensional (1D) tight-binding (TB) chain with a single orbital per lattice site and nearest-neighbor hopping parameter $\gamma$ (note that similar model has been treated numerically in Ref.~\cite{Sayad2015} using finite-length TB chain and nonequilbrium drive different from light). Thus, the matrix elements of the electronic Hamiltonian in the position representation are $H_{n,n+1,\sigma\sigma'} = -\gamma\delta_{\sigma\sigma'}$, and electronic band dispersion is $\varepsilon_\sigma(k)=-2\gamma \cos(ka)$. Here, $\delta_{\sigma\sigma'}$ is the Kronecker delta and $a$ is the lattice constant. The internal magnetic field due to integrated-out electrons $\mathbf{B}^e_n$ in Eq.~\eqref{eq:modifiedLLG}, as given by the integral in Eq.~\eqref{eq:electronB}, vanishes due to the time-reversal symmetry of the electronic Hamiltonian. In turn, the non-Markovian memory kernel $\eta(t,t')\equiv \eta_{11}(t,t')$ due to the same integrated-out electrons is nonzero and obtained by numerically computing Eq.~\eqref{eq:nonmarkovianKernel} for specific time profiles of the vector potential $\mathbf A(t)$ of applied light. Figure~\ref{fig:nmkernels} shows the structure of $\eta(t,t')$ in $t$-$t'$ plane for different fsLPs whose vector potential $\mathbf{A}(t) = z \exp\left(-\frac{t^2}{2\lambda^2}\right) \cos(\omega_L t)\mathbf e _x$ contains a Gaussian envelope and where $z=eA_{\rm max}/a\hbar$ is the dimensionless amplitude~\cite{Bajpai2019a} of the vector potential; $\lambda$ determines the duration of fsLP; and $\omega_L$ is its central frequency. In the absence of light, $z=0$, the non-Markovian memory kernel has a finite value only close to the time diagonal $t= t'$ [Fig.~\ref{fig:nmkernels}(a)], meaning that localized spin is not influenced by its orientation at previous times. This is because the characteristic timescale for the dynamics of electrons in the absence of light is governed by their Fermi energy and, therefore, is much shorter than that of slow classical localized spins. In other words, in the absence of light, the non-Markovian memory kernel is well-approximated by a first order truncation of its series in powers of $t-t'$, reducing to the nonlocal-in-space but local-in-time damping tensor found in Refs.~\cite{ReyesOsorio2024, ReyesOsorio2023}. Conversely, introducing light-electron coupling into this system makes the memory kernel fully non-Markovian in Figs.~\ref{fig:nmkernels}(b)--(d), i.e., it is finite away from the time diagonal $t'<t$ extending to all previous times at which fsLP is operative. Furthermore, increasing amplitude of fsLP from $z=1$ to $z=5$ leads to self-similar fractal structure of $\eta(t,t')$ when plotted in the $t$-$t'$ plane~\cite{Theiler1990}, for which we obtain (via the box-counting method~\cite{Theiler1990}) noninteger dimension $d=1.2$ [Figs.~\ref{fig:nmkernels}(c) and~\ref{fig:nmkernels}(d)]. Note that increasing the duration of fsLP from Fig.~\ref{fig:nmkernels}(c) to~\ref{fig:nmkernels}(d), at the same $z$, makes the self-similarity of unraveled fractals even more conspicuous. 

The properties of $\eta(t,t')$ unraveled by Fig.~\ref{fig:nmkernels} directly translate into the physics of spin dynamics driven by a train of fsLPs, as displayed in Fig.~\ref{fig:onespin_nm}. For example, for low amplitude $z=1$ of fsLPs, we find additional light-induced peaks in the fast Fourier transform (FFT) amplitude spectra  of $S^x(t)$ trajectories [Fig.~\ref{fig:onespin_nm}(a) and~\ref{fig:onespin_nm}(c)]. Otherwise, in the absence of light, FFT amplitude spectra exhibit only a single peak encoding the precession of spin around the $z$-axis due to applied external static magnetic field along that axis. In addition, for sufficiently intense fsLP, $z=5$ [Fig.~\ref{fig:onespin_nm}(b) and~\ref{fig:onespin_nm}(d)], we observe ultrafast \textit{switching} of spin orientation from positive to negative $z$-axis. Such magnetization switching via optical (instead of traditional current injection driven~\cite{Ralph2008}) STT has been amply explored experimentally~\cite{Nemec2012,Zhang2022a,Huebner2024} and for technological applications~\cite{Kimel2019,Huebner2024}. However, its microscopic understanding, and thereby ability to control it, is lacking, as the STT term introduced into the LLG equation in prior attempts to model such an effect was simplistic and justified by intuitive reasoning~\cite{Nemec2012,FernandezRossier2004a,Nunez2004}.

When the amplitude of fsLP is $z \lesssim 1$, the memory effects in the extended LLG Eq.~\eqref{eq:modifiedLLG} are diminished [Fig.~\ref{fig:nmkernels}(b)]. This makes it possible to approximate the non-Markovian kernel $\eta_{nn'}(t,t')$ by a second order truncation of its Taylor series in powers of $t-t'$, thereby leading to time-local iLLG Eq.~\eqref{eq:inertia} with an optically induced magnetic inertia term. Note that in Eq.~\eqref{eq:inertia} we emphasize that the magnitudes of \textit{both} nonlocal damping and magnetic inertia are governed by \textit{time-dependent} prefactors $\lambda_{nn'}(t)$ and $I_{nn'}(t)$, respectively. In the case of light-driven itinerant electrons within 1D TB chain, we compute these functions analytically as
\begin{equation}\label{eq:parameters}
    \frac{\lambda_{nn'}(t)}{\eta_0} = \frac{I_{nn'}(t)}{\eta_1}= \left[\frac{1}{1- |\mathbf{A}(t)|^2} + \cos(2k_F|n-n'|) \right],
\end{equation}
where $\eta_0$ and $\eta_1$ are parameters proportional to $J_{sd}^2$.

To gain insight into the effect of optically induced magnetic inertia, we apply Eq.~\eqref{eq:inertia} to our second simple model in which 30 localized spins,  $\mathbf{S}_n(t)$ where $n=1$--$30$, interact via $sd$ exchange of strength $J_{sd}$ with spins of itinerant electrons hosted by an infinite 1D TB chain. The localized spins also interact with each other via Heisenberg exchange $J$, as described by their classical Hamiltonian 
\begin{equation}\label{eq:spinHam}
    \mathcal{H} = -\sum_n \big[J\mathbf{S}_n\cdot \mathbf{S}_{n+1} + K(S^z_n)^2\big],
\end{equation}
where $K=J/2$ introduces magnetic anisotropy along the $z$-axis. The itinerant electrons are driven by continuous-wave (CW) light described by vector potential \mbox{$\mathbf{A}(t)=0.5\sin(\omega_L t) \mathbf{e}_x$}, and light-induced parameters are set as $\eta_0=0.01$ and $\eta_1=0.08\hbar/J$. The optically excited magnon spectrum in Fig.~\eqref{fig:spectra} is extracted~\cite{Kim2010,Moreels2024} by performing FFT of $S^x_n(t)$ for $n=15$ in the middle of the chain. Let us recall that the LLG equation describes magnons as classical spin waves~\cite{Pirro2021,Skubic2008,GarciaGaitan2025,Kim2010} arising due to collective precessional
motion of localized spins $\mathbf{S}_n(t)$. The initial orientation  $\mathbf{S}_n(t=0)$ is randomized in accord with a thermal distribution, and FFT amplitudes are averaged over 100 such configurations. Note that evolving such an initial state in the absence of CW light produces incoherent (or thermal) magnons~\cite{De2024,Pirro2021} over a continuous band $J\leq \hbar\omega \lesssim 4J$ of frequencies [enclosed  within the left gray box in Fig.~\ref{fig:spectra}(a), as a guide to the eye]. Once CW light is applied, the FFT amplitude spectrum reveals the emergence of four {\em sharp peaks} of coherent~\cite{De2024,Pirro2021} magnons in Fig.~\ref{fig:spectra}(a). Their frequencies,  $\omega_\pm^{(1)} = \omega_L \pm 0.23/\eta_1$ and $\omega_\pm^{(3)} = 3\omega_L \pm 0.23/\eta_1$, are governed by the frequency of light $\omega_L$ and the strength of magnetic inertia [determining parameter $\eta_1$ via Eq.~\eqref{eq:parameters}]. The recent experimental efforts~\cite{Hortensius2021, Wang2023a} in optical excitation of magnons have been intensely focused on understanding how to generate them at arbitrary wavevector $\mathbf k$, instead of usually achieved $\mathbf{k}=0$ magnons. We unravel range of $\mathbf{k}$ vectors of coherent magnon peaks in Fig.~\ref{fig:spectra} using Fig.~\ref{fig:wavevector} within End Matter. Such optically induced magnetic inertia is also responsible for an \textit{additional} band of incoherent magnons at frequencies $6J\lesssim \hbar\omega \lesssim 10J$, which is associated with nutation in $\mathbf{S}_n(t)$ trajectories [enclosed within the right gray box in Fig.~\ref{fig:spectra}(a)]. While the band of nutation-generated magnons has been discussed~\cite{Kikuchi2015, Lomonosov2021, Mondal2022, Daquino2023, Titov2022} for magnetic inertia terms derived due to mechanisms not involving light, for which $I_{nn'}$ is a time- and spatially independent constant, optically induced magnetic inertia generates two sharp peaks of coherent magnons outside either precessional thermal or nutational bands of incoherent magnons in Fig.~\ref{fig:spectra}(a). The same result is replicated (dashed red line) in Fig.~\ref{fig:spectra}(b), which plots the spectra of optically excited magnons for a range $\omega_L$ with coherent magnon frequencies [yellow traces in Fig.~\ref{fig:spectra}(b)] depending linearly on it.

\textit{Conclusions and Outlook}---We demonstrate a field-theoretic derivation of an {\em extended} LLG equation which describes opto-magnetic coupling within light-driven magnets as an effect mediated by fast photoexcited itinerant electrons that are integrated out within the functional integral of SKFT to arrive at LLG-type equation for slow localized spins. This approach displaces the need for various  phenomenological terms and physical pictures that have been invoked~\cite{Kampfrath2010,Tzschaschel2017,Rongione2023} to describe direct coupling of light to local magnetization, or STT of photocurrent~\cite{Ulrichs2018, Ritzmann2020, Nemec2012,FernandezRossier2004a,Nunez2004} on local magnetization, by adding such terms into the standard (Markovian) LLG equation~\cite{Landau1935,Galkina2021}.  The simplicity of phenomenological terms {\em cannot} replicate the complexity of rigorously derived  Eqs.~\eqref{eq:modifiedLLG} and~\eqref{eq:inertia} as our {\em central}  results. This is because  Eq.~\eqref{eq:modifiedLLG} contains a \textit{non-Markovian} memory kernel expressed in terms of Keldysh GF [Eq.~\eqref{eq:nonmarkovianKernel}], whose expansion (for sufficiently low intensity of light) in Eq.~\eqref{eq:inertia} reveals {\em optically induced magnetic inertia}.  Such inertia term, which is distinct from previously derived~\cite{Mondal2017, Bajpai2019, Quarenta2024} inertia terms by having a time-dependent prefactor due to light, is shown to generate sharp peaks of coherent magnons governed by the frequency of light and properties of electron--localized-spin interaction. Thus, we anticipate that coding our Eqs.~\eqref{eq:modifiedLLG} and~\eqref{eq:inertia} into LLG-based  simulation packages~\cite{Evans2014, Skubic2008, Vansteenkiste2014,Moreels2024} will offer an accurate tool for understanding and controlling optical magnetization switching~\cite{Nemec2012,Kimel2019,Huebner2024} or excitation of magnons~\cite{Afanasiev2021, Allington2025, Hortensius2021, Zhang2020a, Diederich2025} in recent experiments, as also intensely pursued for applications in magnonics~\cite{Pirro2021,Flebus2024} and quantum information~\cite{Gish2024, Lauk2020, Hisatomi2016}. Finally, we note that our theory is not directly applicable to fsLP interacting with electrons having a gap in their energy spectrum, as exemplified by magnetically ordered Mott insulators~\cite{GarciaGaitan2025a,Metzger2025,Rongione2023} or semiconductors~\cite{Murakami2020,Seifert2022a}, where often used~\cite{Satoh2010,Rongione2023,Tzschaschel2017,Allington2025} laser light of subgap frequency never photoexcites itinerant electrons we consider in our theory. But even in those cases, light should be coupled to electrons via  an appropriate quantum many-body (such as Hubbard~\cite{GarciaGaitan2025a,Murakami2020,Murakami2025a,Seifert2022a})  Hamiltonian, instead~\cite{Metzger2025} of   assuming ``off-resonant''~\cite{Rongione2023} direct coupling of light to local magnetization. 


\textit{Acknowledgments}---This work was supported by the US Department of Energy (DOE) Grant No. DE-SC0026068.

%

\section{End Matter}

\subsection{Models and SKFT-based Methodology}
Our starting point is a fully quantum many-body Hamiltonian describing the interaction of itinerant electrons and localized quantum spins, $\hat H = \hat H_e + \hat H_{sd} +  \hat H_S$. The first term, $\hat H_e$, is a general Hamiltonian for electrons on a TB lattice subjected to an external electromagnetic field
\begin{equation}\label{eq:hamiltonian_pos}
    \hat H_e = \sum \hat c^\dagger_{n\sigma} H_{nn'}^{\sigma\sigma'} \hat c_{n'\sigma'} \, e^{i(\mathbf{r}_{n} - \mathbf{r}_{n'})\cdot e \mathbf{A}(t)}.
\end{equation}
Here, $\hbar=1$; $\hat c^\dagger_{n\sigma} \ (\hat c_{n\sigma})$ creates (annihilates) electrons with spin $\sigma$ on lattice site $n$ and position vector $\mathbf{r}_{n}$; $H_{nn'}^{\sigma\sigma'}$ is a matrix element of the electronic Hamiltonian; $e$ is the electron charge; and $\mathbf{A}(t)$ is spatially uniform, but time-varying vector potential in the Coulomb gauge which couples to the Hamiltonian via the Peierls phase~\cite{Eckstein2020, Panati2003}. The momentum representation of the Hamiltonian in Eq.~\eqref{eq:hamiltonian_pos} is given by
\begin{equation}\label{eq:hamiltonian_mom}
    \hat H_e = \sum \varepsilon_\sigma(\mathbf{k}-e\mathbf{A}) \hat \psi^\dagger_{\mathbf{k}\sigma} \hat \psi_{\mathbf{k}\sigma},
\end{equation}
where $\varepsilon_\sigma(\mathbf{k})$ is the energy-momentum dispersion; \mbox{$\hat \psi_{\mathbf{k}\sigma} = \sum_{n \sigma'} U_{\sigma\sigma'}(\mathbf{k}) e^{i\mathbf{k\cdot r}_n} \hat c_{n \sigma'}$} annihilates an electron with momentum $\mathbf{k}$ and spin $\sigma$; and $U_{\sigma\sigma'}(\mathbf{k})$ is the matrix that diagonalizes Eq.~\eqref{eq:hamiltonian_pos} in the basis of Bloch states. We assume that itinerant electrons to which light couples directly in Eqs.~\eqref{eq:hamiltonian_pos} and~\eqref{eq:hamiltonian_mom} are either intrinsic to a ferromagnetic metal irradiated by light or provided by a normal metal layer attached~\cite{Bertelli2021,ReyesOsorio2023} to a ferromagnetic insulator. The Hamiltonian commutes with itself at different times, allowing for straightforward determination of the Keldysh GFs~\cite{Stefanucci2013, Turkowski2005}
\begin{subequations}
\begin{align}
    G^R_{\mathbf{k}\sigma} &= -i\theta(t-t')e^{-i\int_{t'}^t \! ds \, \varepsilon_\sigma(\mathbf{k}-e\mathbf{A}(s))}, \\
    G^A_{\mathbf{k}\sigma} &= i\theta(t'-t)e^{-i\int_{t'}^t \! ds \, \varepsilon_\sigma(\mathbf{k}-e\mathbf{A}(s))}, \\
    G^K_{\mathbf{k}\sigma} &= -i\big[1-2f\big(\varepsilon_\sigma(\mathbf{k})\big)\big]e^{-i\int_{t'}^t \! ds \, \varepsilon_\sigma(\mathbf{k}-e\mathbf{A}(s))},
\end{align}
\end{subequations}
where superscripts $R/A/K$ refer to the retarded, advanced, and Keldysh components of the GF, respectively, and $f(\varepsilon)$ is the Fermi-Dirac distribution. The electronic spin density operator, $\hat{\mathbf{s}}_{n}=\sum\boldsymbol{\sigma}_{\sigma\sigma'} \hat c^\dagger_{n \sigma} \hat c_{n \sigma'}$ where $\boldsymbol{\sigma}$ is the vector of the Pauli matrices,  couples to the localized spin operator $\hat{\mathbf{S}}_{n}$ at the site $n$ via $sd$ exchange of magnitude $J_{sd}$, as described by \mbox{$\hat H_{sd} = -J_{sd}\sum_{n} \hat{\mathbf{S}}_{n}\cdot \hat{\mathbf{s}}_{n}$}. Finally, $\hat H_S$ is the Heisenberg-type Hamiltonian describing interaction between  $\hat{\mathbf{S}}_n$ operators [such as Eq.~\eqref{eq:spinHam}, but with $\mathbf{S}_n \mapsto \hat{\mathbf{S}}_n$]. 

Constructing a quantum field-theoretic description of the same system requires real fields $\mathbf{S}_n$ for the localized spins, which are obtained as the expectation value of spin operators in the spin coherent states~\cite{Altland2010}, \mbox{$\braket{\theta,\phi| \hat{\mathbf{S}}_n | \theta,\phi} = \mathbf{S}_n$}. In addition, for describing electrons we need Grassmann fields $\psi_\mathbf{k}$ as the eigenvalues of the fermionic coherent states, \mbox{$\hat\psi_\mathbf{k} \ket{\psi_\mathbf{k}} = \psi_\mathbf{k}\ket{\psi_\mathbf{k}}$}~\cite{Kamenev2023}. The SKFT action \mbox{$S=S_S + S_e$} corresponding to the Hamiltonian $\hat H$ is the sum of two terms
\begin{subequations}\label{eq:action}
	\begin{align}
        S_S &= \int\! dt \sum_n \mathbf S_{n}^q\cdot \left(\partial_t \mathbf S_{n}^c \times \mathbf{S}_{n}^c + \mathbf{B}^{\rm eff}[\mathbf{S}^c_{n}]\right), \label{eq:actionS}\\
        S_e &= \int \! dtdt' \sum_{\mathbf k \sigma} \bar{\boldsymbol{\psi}}_{\mathbf{k}\sigma}(t) \check G_{\mathbf{k}\sigma}(t,t') \boldsymbol\psi_{\mathbf{k}\sigma}(t') \nonumber \label{eq:actionE}\\
        & +\int\! dt \sum_{n}\sum_{\alpha\sigma\sigma'} J_{sd} \bar{\boldsymbol{\psi}}_{n\sigma} \check{S}^\alpha_{n}\sigma^\alpha_{\sigma\sigma'}\boldsymbol{\psi}_{n\sigma}, 
	\end{align}
\end{subequations}
where $S_S$ and $S_e$ are the actions of the localized spins and itinerant electrons, respectively; $S_e$ includes the contribution from the $sd$ exchange; and $\mathbf{B}^{\rm eff}[\mathbf{S}^c_{n}] = -\partial\mathcal H/\partial \mathbf{S}\big|_{\mathbf{S}^c_n}$ is the effective magnetic field due to localized spins described by the classical Hamiltonian $\mathcal H$~\cite{Evans2014,Skubic2008}, such as Eq.~\eqref{eq:spinHam} we employ in this study (we set the Bohr magneton $\mu_B=1$). The superscripts $c,q$ refer to the classical and quantum components of the localized spin field given by the so-called Keldysh rotation~\cite{Kamenev2023} $\mathbf S^\pm_n = \mathbf S^c_n \pm \frac{1}{2}\mathbf S^q_n$, where $\mathbf{S}^\pm_n$ are the fields on the forward and backward segments of the Keldysh closed time contour~\cite{Kamenev2023, Stefanucci2013}. Similarly, $\boldsymbol{\psi}_\mathbf{k}=(\psi^1_\mathbf{k}, \psi^2_\mathbf{k})^T$, where $\psi^\pm_\mathbf{k} = \frac{1}{\sqrt{2}}(\psi^1_\mathbf{k} \pm \psi^2_\mathbf{k})$ and $\bar\psi^\pm_\mathbf{k} = \frac{1}{\sqrt{2}}(\bar\psi^2_\mathbf{k} \pm \bar\psi^1_\mathbf{k})$. We additionally use $\check O$ notation for $2\times 2$ matrices, such as
	\begin{equation}
		\check G_{\mathbf k \sigma} = \begin{pmatrix}
			G^R_{\mathbf k \sigma} & G^K_{\mathbf k \sigma} \\
			0     & G^A_{\mathbf k \sigma}
		\end{pmatrix}, \quad 
		\check{S}_{n}^\alpha = \begin{pmatrix}
			S^{c,\alpha}_n  & \frac{1}{2}S^{q,\alpha}_n \\
			\frac{1}{2}S^{q,\alpha}_n    &  S^{c,\alpha}_n 
		\end{pmatrix},
	\end{equation}
in the so-called Keldysh space~\cite{Kamenev2023}.

\begin{figure}
    \centering
    \includegraphics[width=\columnwidth]{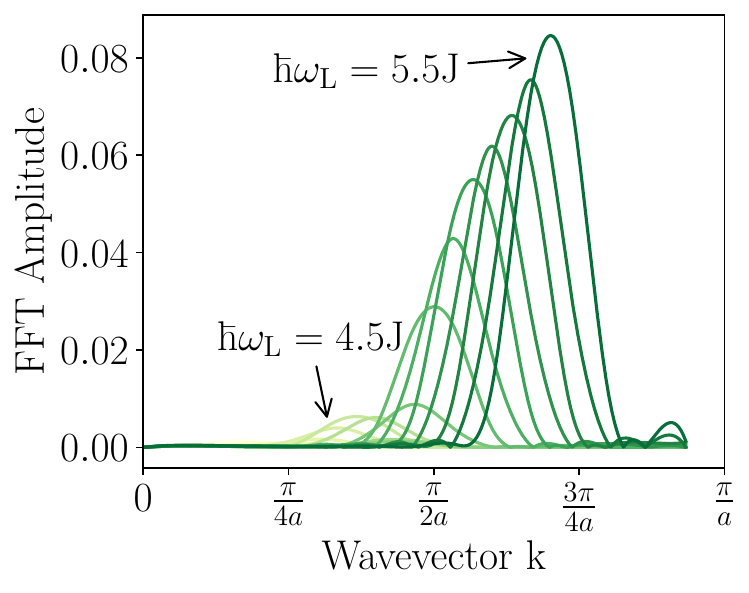}
    \caption{Wavevector of magnons in a chain of 30 localized spins externally driven by CW light of varying frequency $\omega_L$ [same setup employed in Fig.~\ref{fig:spectra}] extracted from spatial FFT amplitude spectrum of $\mathbf{S}_n(t)$, on the proviso that chosen time $t$ is late enough so that only the long-lived light-generated coherent magnons [sharp peaks in Fig.~\ref{fig:spectra}(a)] contribute to the spectrum. Darker curves correspond to higher $\omega_L$.}
    \label{fig:wavevector}
\end{figure}

The electronic action $S_e$ in Eq.~\eqref{eq:actionE} is integrated out perturbatively to second order in the $J_{sd}$ coupling~\cite{ReyesOsorio2024}, thereby yielding an effective action for localized spins \textit{only}, $S_S^{\rm eff} = S_S + S'$, where $S'$ is given by
\begin{equation}\label{eq:actionEff}
    S' =  \int\! dt \sum_{n} \mathbf{S}^q_{n}\cdot \Big[ \mathbf{B}_{n}^{\rm e} + \int\! dt' \sum_{n'} \eta_{nn'}(t,t')\mathbf{S}_{n'}^c (t') \Big].
\end{equation}
Here a stochastic noise due to terms quadratic in the quantum fields has been neglected in order to focus on the deterministic limit~\cite{ReyesOsorio2024, Verstraten2023}. Within this action, localized spins are exposed to light indirectly via light driving electrons out of equilibrium, so that their integration out introduces a local magnetic field into Eq.~\eqref{eq:modifiedLLG}
\begin{equation}\label{eq:electronB} 
\mathbf{B}_{n}^e = \frac{-iJ_{sd}}{2}\int_{\rm BZ}\!\frac{d\mathbf{k}}{(2\pi)^d} \, {\rm Tr}_{\rm spin}\Big[\underline G_\mathbf{k}^K(t,t) \boldsymbol{\sigma} \Big], 
\end{equation}
as well as non-Markovian memory kernel, $\eta_{nn'}^{\alpha\beta}(t,t')$. This kernel is also nonlocal and generally given by the tensor
\begin{align}
\eta_{nn'}^{\alpha\beta}(t,t') &= \frac{-iJ_{sd}^2}{2}\theta(t-t') \int_{\rm BZ}\!\frac{d\mathbf{k}d\mathbf{q}}{(2\pi)^{2d}} \, e^{i(\mathbf{k-q})\cdot (\mathbf{r}_n - \mathbf{r}_{n'})} \nonumber \\
&\times {\rm Tr}_{\rm spin} \Big[  \sigma^\alpha \underline G_\mathbf{k}^R(t,t') \sigma^\beta \underline  G_\mathbf{q}^K(t',t)\Big] + \textrm{H.c.} , \label{eq:nonmarkovianKernel} 
\end{align}
where $\underline G_\mathbf{k} \equiv U^\dagger (\mathbf k) G_\mathbf{k}  U(\mathbf{k})$; the integrals $\int_{\rm BZ}$ are over the first Brillouin zone; the partial trace ${\rm Tr_{spin}}$ is over spin space only; and $\theta(t-t')$ is the unit step function. Minimizing the effective action, via its functional derivative being $\delta S_S^{\rm eff}/\delta \mathbf{S}^q_n=0$, yields the non-Markovian extended LLG Eq.~\eqref{eq:modifiedLLG} for classical components of spin fields $\mathbf{S}^c_n$ [note that for simplicity we drop the superscript $c$, using $\mathbf{S}_n \equiv \mathbf{S}^c_n$ in Eqs.~\eqref{eq:modifiedLLG} and~\eqref{eq:inertia}].

For numerically integrating the iLLG Eq.~\eqref{eq:inertia}, additional variables $\mathbf{L}_n = \mathbf{S}_n\times \partial_t \mathbf{S}_n$ are typically employed~\cite{Daquino2024} to convert it into a system of first-order differential equations. However, the nonlocality of our SKFT-derived iLLG Eq.~\eqref{eq:inertia} suggests the necessity of many more such variables---such as, $\mathbf{L}_{nn'} = \mathbf{S}_n\times \partial_t \mathbf{S}_{n'}$ for all pairs of sites $n$ and $n'$---thereby making its numerical integration quite challenging. Thus, for Fig.~\ref{fig:spectra} we simplify numerics by spatially averaging the nonlocal damping and inertia terms, i.e., $\lambda_{nn'}(t) \mapsto \lambda_{nn}(t)\delta_{nn'}$ and $I_{nn'}(t)\mapsto I_{nn}(t)\delta_{nn'}$. Note that in both Eqs.~\eqref{eq:modifiedLLG} and~\eqref{eq:inertia}, we add a conventional local-in-time Gilbert damping term~\cite{Gilbert2004, Saslow2009, Evans2014} whose magnitude is governed by $\alpha_G$ as material-dependent constant that is measured experimentally~\cite{Weindler2014} or computed by including spin-orbit coupling~\cite{Guimaraes2019,Thonig2017} or phonon~\cite{Verstraten2023, Quarenta2024} or both~\cite{Liu2011f} effects. In all calculations in Figs.~\ref{fig:onespin_nm} and~\ref{fig:spectra}, we use $\alpha_G=0.003$.  

\subsection{Wavevector of coherent light-generated magnons}

In order to determine the wavevector $\mathbf{k}$ of the light-generated coherent magnons, which correspond to the sharp peaks in Fig.~\ref{fig:spectra}(a) or yellow traces in Fig.~\ref{fig:spectra}(b), we apply a FFT in space to the setup of our second simple model, the chain of 30 localized spins $\mathbf{S}_n(t)$ driven by CW light. The time $t$ is chosen to be late enough in the evolution so that most magnons have decayed except for the light-generated coherent ones. The FFT amplitude spectrum is plotted in Fig.~\ref{fig:wavevector}. The wavevector $\mathbf{k}$ increases with the frequency of light $\omega_L$, starting from $|\mathbf{k}|\approx\pi/4a$. 

\end{document}